\documentclass[a4paper,11pt]{article}
\usepackage{pos}
\usepackage{lineno}
\title{The Real Time Analysis framework of the Cherenkov Telescope Array's Large-Sized Telescope}
 \ShortTitle{Real Time Analysis LST-CTA}

\author*[a]{Sami Caroff}
\author[a]{Pierre Aubert}
\author[a]{Enrique Garcia}
\author[a]{Gilles Maurin}
\author[a]{Vincent Pollet}
\author[a]{Thomas Vuillaume}

\affiliation[a]{Univ. Savoie Mont Blanc, CNRS, Laboratoire d’Annecy de
Physique des Particules - IN2P3,\\
  74000 Annecy, France}

\onbehalf{on behalf of the CTA-LST Project} 

\emailAdd{sami.caroff@lapp.in2p3.fr}
\abstract{The Large-Sized Telescopes (LSTs) of the Cherenkov Telescope Array Observatory (CTAO) will play a crucial role in the study of transient gamma-ray sources, such as gamma-ray bursts and flaring active galactic nuclei. The low energy threshold of LSTs makes them particularly well suited for the detection of these phenomena. The ability to detect and analyze gamma-ray transients in real-time is essential for quickly identifying and studying these rare and fleeting events. In this conference, we will present recent advances in the real-time analysis of data from the LST-1, the first prototype of LST located in the Canary island of La Palma. We will discuss in particular the development of new algorithms for event reconstruction and background rejection. These advances will enable rapid identification and follow-up observation of transient gamma-ray sources, making the LST-1 a powerful tool for the study of the dynamic universe. The implementation of this framework in the future Array Control and Data Acquisition System (ACADA) of CTAO will be discussed as well, based on the experience with LST.}

\ConferenceLogo{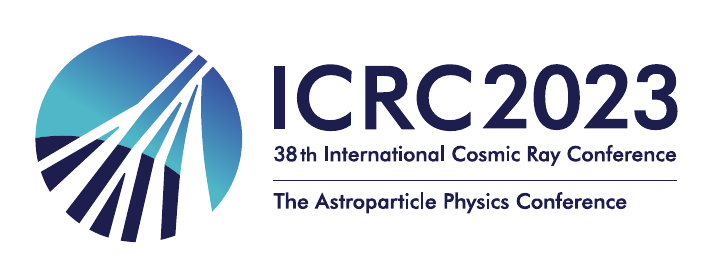}

\FullConference{%
38th International Cosmic Ray Conference (ICRC2023)\\
  26 July - 3 August, 2023\\
  Nagoya, Japan}

\begin{document}
\maketitle

\section{Introduction}

Gamma-ray astronomy aims to study the cosmic particle accelerators of the universe, through the analysis of their high-energy electromagnetic emissions. It covers various types of sources, from Galactic ones such as supernovae remnants and pulsar wind nebulae to extragalactic ones like active galactic nuclei. Recently, the first detections of the  afterglow radiations from gamma-ray bursts \citep{2019Natur.575..464A,2019Natur.575..459M,2021Sci...372.1081H} were achieved at very high energies (VHE, $E > 100$ GeV), opening a new research area around transient phenomena. Moreover, recent observations of the electromagnetic counterpart of gravitational wave emitters have motivated a fast development of the field. 

The Cherenkov Telescope Array Observatory (CTAO) represents the next generation of ground-based observatories. In October 2018, the first prototype of the Large-Sized Telescope, called LST-1, was installed and inaugurated at the Roque de los Muchachos Observatory (ORM). It has been actively collecting data since November 2019. The LST-1 boasts a 23-meters diameter reflector dish, a lightweight mechanical structure, and a rapid repositioning system. Its primary purpose is to detect low-energy gamma rays, above approximately $20$ GeV, and to enable swift follow-up observations of transient events.
%It will be located on two separate sites: a southern site situated in the Atacama Desert in Chile and a northern site at the Roque de los Muchachos observatory (ORM) on the island of La Palma in the Canary Islands, Spain. The southern site will house 51 telescopes of two different sizes, namely Medium-Sized Telescopes and Small-Sized Telescopes. On the other hand, the northern site will feature 13 telescopes of two different sizes, namely Large-Sized Telescopes and Medium-Sized Telescopes.
Efficiently monitoring alerts and observing transient events is only possible with a robust and fast (real-time) analysis system, which processes data immediately after they are collected during observations and provides timely feedback to operators to initiate appropriate decisions and generate alerts. The analysis process can be divided into two sequential stages: first, the reconstruction, characterization, and selection of gamma rays from the recorded images, and second, the search for sources within the telescope's field of view. 
   
This proceeding only covers the first part which is the real-time analysis (RTA) of the LST-1, prototype of the CTAO online-reconstruction. Section \ref{sec:Analysis} describes the global architecture and the technical options adopted while section \ref{perfomance} exposes the performance and results obtained on Markarian 421 data. 

\section{Data Analysis}
\label{sec:Analysis}
The initial stage of the analysis process, known as reconstruction, focuses on processing the uncalibrated raw data (R0) to derive the parameters of each recorded event (DL3-data). These parameters include the energy, direction, and a gamma-ray like classification score (called gammaness). The LST-1 telescope R0 data consists of a sequences of 40 images, each containing 1855 pixels, for every recorded event. In its nominal mode, the telescope operates at a rate of $\sim$10 kHz, generating a data flow of $\sim$3 GB/s of the raw data but the rate can go up to 15 kHz in case of extreme astrophysical events. Looking ahead to the future CTA observatory, the expected data flow will be approximately 17 GB/s for the northern site and 27 GB/s for the southern site. 
%However, it is important to note that these data volumes are not directly proportional to the number of telescopes since a central trigger system will selectively choose stereoscopic events for analysis.

%The reconstruction process involves three interfaces : 
% \begin{itemize}
% \item it receives the raw R0 data in stream in the Protocol Buffer format \citep{protobuf} via ZeroMQ \citep{refZeroMQ}, the open-source universal Library messaging,
% \item it generates DL3-data in Flexible Image Transport System format (FITS), 
% \item it receive telescope pointing direction in celestial coordinates from the Telescope Control software of the LST1.
% \end{itemize}

The historical Hillas method \cite{refHillasPaper} serves as the reference reconstruction approach, renowned for its simplicity and robustness. Its modern version comprises three consecutive stages, implemented in three corresponding software blocks:
    \begin{itemize}
        \item The \textbf{R0->DL1} stage includes data calibration, images aggregation and cleaning for each event. It also encompasses the extraction of Hillas and timing parameters (as defined in \cite{LST_Crab}). This stage significantly reduces the data volume from several GB/s to less than 150 MB/s. Most of the output data consists of the final calibrated images used for monitoring the data acquisition (DAQ), while the Hillas parameters correpond to only a few MB/s.
        \item The \textbf{DL1->DL2} stage focuses on evaluating the energy, direction, and gammaness. The input and output data sizes are approximately the same, amounting to a few MB/s.
        \item \textbf{DL2->DL3} stage corresponds to the selection of gamma-like events, further reducing the data flow to a few kB/s. 
    \end{itemize}

\subsection{Software Architecture}
To ensure seamless data acquisition, the RTA is performed on dedicated servers, with communication between the DAQ system and the analysis servers established through the network. At least two servers are required to receive data from the LST-1. R0 data is streamed in the Protocol Buffer format\footnote{\url{https://developers.google.com/protocol-buffers/docs/overview}} handled by ZeroMQ\footnote{\url{https://zeromq.org}}, an open-source universal messaging library.

The Slurm\footnote{\url{https://slurm.schedmd.com/documentation.html}} workload manager plays a vital role in resources and processes management. This open-source, fault-tolerant, and highly scalable cluster management and job scheduling system is specifically designed for both large and small Linux clusters. 
%It ensures optimal utilization of available resources.

On each server, four processes of R0->DL1, each with two threads, continuously run to process the data stream. Each thread handles an approximate event rate of 1.25 kHz, resulting in a total processing capacity of 10 kHz under nominal operation. Evaluation of the events produced by Car flash\footnote{Car flash events are induced by the illuminations of car passing by the road near the telescope.} indicates that the maximum rate of 15 kHz (1.875 kHz) provided by the LST-1 is supported by the analysis servers. A short buffer of 100 events, equivalent to 80 ms of data, is used.%, and if the buffer becomes full, events are discarded to maintain data flow continuity.

Following the R0->DL1 computation, HDF5\footnote{\url{https://support.hdfgroup.org/HDF5/}}\footnote{\url{https://gitlab.in2p3.fr/CTA-LAPP/PHOENIX_LIBS/PhoenixHDF5/}} files containing $20\,000$ DL1 events are generated. For each new file, the DL1->DL2->DL3 chain is immediately executed using the capabilities of the Slurm manager. 
%Specifically, a process handles the DL1->DL2 task, generating a DL2-event file in HDF5 format. Subsequently, the DL2->DL3 process is initiated, generating a reconstructed DL3-event file in FITS format. 
%These files can then be directly utilized by the high-level analysis for Science Alert Generation. 
%These steps are efficiently performed by the offline LST pipeline using the lstchain software \cite{cite_lstchain}. The HDF5 files serve as buffers, striking a balance between computation and memory resource usage while maintaining the overall execution time of the pipeline. 
%Additionally, hosting the files on a shared filesystem enables other systems to access the produced data from any machine.

\subsection{Optimization of data processing}

%Efficiently processing the high-volume raw data requires careful optimization of the calculation time for each software component and the interfaces connecting them. Figure \ref{figR0DL1} illustrates the sequential stages involved in the R0->DL1 process. These stages include image calibration, where the 40 images of each event are calibrated. The relevant images are then selected and integrated, followed by image cleaning and the extraction of Hillas and temporal parameters. The calibrated and integrated images are stored in the DL1 file, along with the reduced parameters.
The RTA includes several stages: image calibration, selection, integration, cleaning, and extraction of Hillas parameters. The data reduction and processing steps are computationally demanding, therefore, the C++ programming language was chosen for its fast execution and optimization capabilities.
%    \begin{figure}
%        \centering
%        \includegraphics[scale=0.8]{Figure/hiperta_r0_dl1_overview.png}
%        \caption{Overview of the R0->DL1 analysis steps for LST-1 prototype.}
%        \label{figR0DL1}
%    \end{figure}
% to the extensive data reduction and the numerous processing steps involved, this process is highly computationally demanding. Hence, the choice of the C++ programming language is natural, as it ensures fast execution and allows for various optimizations discussed in subsequent sections.

%\subsubsection{calibration}

The calibration step consists in converting electronic charge into the corresponding number of photoelectrons detected by each camera pixel by subtracting a baseline (pedestal) and applying a conversion factor (gain) . The calibration algorithm selects the appropriate gains and pedestals obtained with calibration runs. The compiler automatically vectorizes this algorithm for improved performance.

%The calibration step focuses on converting the electronic charge into the corresponding number of photons detected by each camera pixel. While the future data acquisition (DAQ) system will provide calibrated data (R1-data), the current setup transmits raw R0-data. 
%Initially, the electronic components of the LST camera generate a sequence of 40 images for each triggered event. The analog signal from the pixels is digitized into a 12-bit raw signal. 
%To avoid overflow, two channels, namely high gain and low gain, are employed. The high gain channel is approximately one magnitude more sensitive than the low gain channel. 
%Two channels, namely high gain and low gain, are employed. Thus, if the high gain signal exceeds 3500 counts and overflows, the low gain signal is utilized. The appropriate gains and pedestals are selected accordingly from the nearest in time calibration runs produced. This conditions is expressed as a mathematical expression to avoid branching and enable automatic vectorization. Following this, a simple calibration is applied: the electronic pedestal is subtracted from the signal received by each pixel, and the result is multiplied by a conversion factor that converts the electronic signal to photo-electrons. Leveraging memory alignment and continuity, the compiler automatically vectorizes this algorithm for enhanced performance.

%\subsubsection{Pixel-wise charge integration}

Pixel-wise charge integration involves summing the calibrated signals for each pixel using a specific time window. The compiler can optimize this process by effectively vectorizing the integration calculations.

%The integration process involves partially summing the calibrated signals to enhance comparability between pixels and improve the signal-to-noise ratio. Pixel-wise integration is performed using a 7 ns window, which starts 3 ns before the maximum signal of the current pixel and ends 4 ns after it. As a result, each pixel has its own integration boundaries, and the time of the maximum signal is used to generate a time map for the current event. 
%This time map will be utilized in the subsequent reconstruction steps.

%Due to the matrix storage arrangement, where pixels are represented in columns and signal waveforms in rows, the compiler can effectively vectorize this integration process. Since the calculations within each row are independent of one another, the compiler can optimize the code for improved performance.

%\subsubsection{Image cleaning}

Image cleaning is performed to select relevant pixels and remove those which triggered due to background noise. A two-thresholds algorithm (thresholding on pixel and pixel neighbours charge) is applied using a temporary matrix to ensure efficient memory access for neighboring pixels. Computation is performed instead of branching, which improves computational time by enabling vectorization.

%The pixels in the camera predominantly contain noise, such as the background light from the night sky. 
%To select the relevant pixels in the ambiant noise, a two-threshold algorithm is employed, typically using thresholds of 8 and 4 pe. The first threshold determines if the current pixel has a sufficient overall signal, while the second threshold verifies if it has at least one neighboring pixel with a significant brightness.

%Accessing memory at random addresses is considerably slow. To overcome this, a temporary matrix is created to ensure that the neighboring pixels are stored close together in memory. Thanks to this temporary matrix, the compiler can optimize the computation by vectorizing the operations. 
%Accessing the data from neighboring pixels becomes faster. Similar to the calibration process, computations are performed instead of branching, which helps avoid branch misses and significantly reduces computational time.

%\subsubsection{Hillas and time parameters computation}

Hillas and timing parameters are computed to characterize the shape of the cleaned images and extract relevant information. High-level intrinsic functions\footnote{\url{https://gitlab.in2p3.fr/CTA-LAPP/PHOENIX_LIBS/IntrinsicsGenerator/}} are used to optimize reduction operations and barycenter computations, resulting in faster execution compared to automatic optimization methods.

%The parametrization process involves computing high-level parameters based on the cleaned images and time map. Firstly, ellipse parametrization is applied to the cleaned images using first and second-order moments. Secondly, a timing algorithm is employed to determine the direction of the shower in the camera based on the time map generated during the integration process. Finally, the last algorithm calculates the signal intensity along the border of the camera.
%This information provides valuable insights into the shower's trajectory. Finally, the last algorithm calculates the signal intensity along the border of the camera. This evaluation helps assess the likelihood of a shower being truncated within the camera's field of view.
%, following the principles of Hillas parametrization. This step helps characterize the shape of the image and extract relevant information. 

%The aforementioned algorithms primarily involve reduction operations that may not be efficiently optimized by standard compilers. To enhance their performance, high-level intrinsic functions are utilized. These functions serve as the interface between assembly language and C, allowing the compiler to explicitly optimize the vectorization of reduction operations and barycenter computations. By leveraging these functions, the algorithms achieve approximately three times faster execution compared to automatic optimization methods.
Overall, these optimization techniques aim to reduce the computation time and enhance the performance of the data processing pipeline.

\subsection{High level parameters computation and DL3 production}

The computation of the high-level parameters is not a bottleneck in terms of computing time, therefore the offline analysis chain of the LST-1 is used in order to simplify the maintainability of the software. This offline analysis chain is described in \cite{LST_Crab} and only a brief description will be provided in this proceeding. This analysis pipeline is working with a different DL1 format (hereafter called DL1-lstchain) than the CTAO standard one. Thus to use it a conversion step is required in the pipeline. 
%But since this analysis is not aligned with the DL1 type defined by the CTAO observatory, an extra step of DL1 conversion need to be added in the pipeline.

The high-level parameters, disp, energy, and gammaness are obtained using random forests trained on Monte Carlo simulations of gamma-ray events generated with a zenith angle of 20°. Similar results with respect to those of the offline trained random forest \cite{LST_Crab} are obtained: time gradient is the most important feature for disp reconstruction, followed by psi, and length Hillas parameters, while the length is the most important feature for energy. For gammaness classification, features ranking is more equally distributed.
%These simulations are separated in a trained sample used for the training of the random forest and a test sample that it used to derive the Instrumental Response Functions (IRFs). 

The last part of the analysis consists of a gammaness selection and a conversion to a FITS format compliant with gammapy, a high-level analysis software \cite{gammapy}.

%\begin{figure}
%    \centering
%     \includegraphics[scale=0.4]{Figure/Disp.PNG}
%     \includegraphics[scale=0.4]{Figure/Energy.PNG}
%     \includegraphics[scale=0.5]{Figure/classification.PNG}
%    \caption{Overview of the R0->DL1 analysis steps for LST-1 prototype.}
%    \label{figRFranking}
%\end{figure}

\section{Performance}
\label{perfomance}
\subsection{Computing time}

\begin{figure}
    \centering
     \includegraphics[scale=0.4]{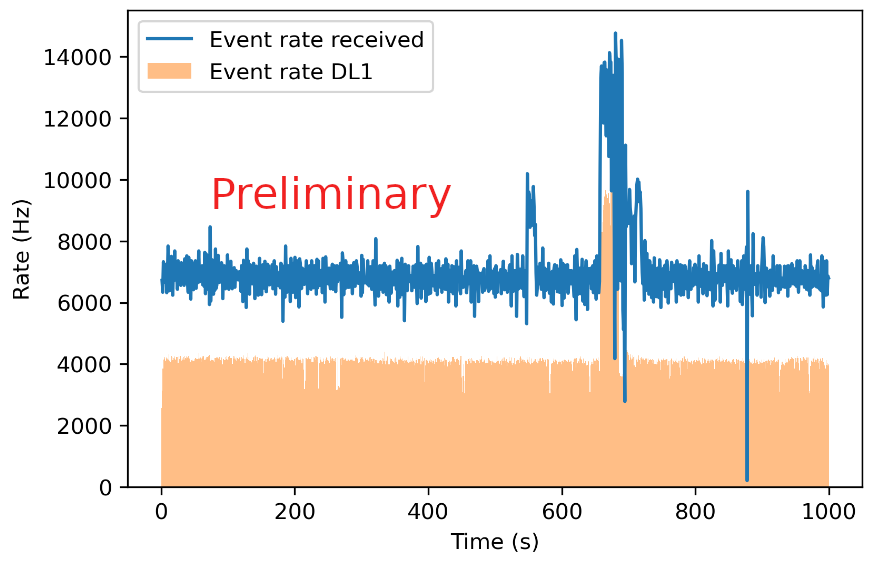}
    \caption{Rate of event received by the R0->DL1 process compared to the rate of events in DL1. The observed difference in normalisation is due to the cleaning step which rejects some of the events.}
    \label{figEventRate}
\end{figure}

As explained in section \ref{sec:Analysis}, the software should support an event rate up to $15$ kHz. Trigger rate peaks are exploited to test this requirement. An example is shown in Figure \ref{figEventRate}: it can be observed that the DL1 reconstruction was not affected by an event rate peak of $14$ kHz. 

The typical time for the production of the different analysis steps, is estimated using an observation run with an average rate of $\sim 8$ kHz: 
\begin{itemize}
\item R0->DL1 : $30$ s. Taking that the effective rate per thread before cleaning is $1$ kHz, and that $20$k events need to be accumulated to create a DL1 file, the analysis time can be approximated as $10$ seconds while the $20$ remaining seconds are spend to accumulate events. The offline chain for the DL1 production is running at $0.03$ s/event while the RTA achieve a computing speed of $0.5$ ms/event.
\item DL1->DL1 lstchain : $12$ s are used for this step, which will disappear when lstchain data format aligns with CTAO's in future releases. 
\item DL1 lstchain->DL2->DL3 : $14$ + $10$ s.
\end{itemize}
In total, $66$ seconds are needed to produce a DL3 from $20$k R0 events. It is important to note that this time scales linearly with the number of events per file. %, and that $30\%$ of this time is accumulation of events coming from the camera which is trigger rate dependent. 

\subsection{Reconstruction}
%These data was taken between May $13$ $2023$ and May $29$ $2023$.
The reconstruction performance is tested using real time data obtained from observations of Markarian 421. They consist of $14$ observations selected according to the weather on site, variability of event rate at the DL3 level and zenith angle lower than $30$°, amounting to $4.2$ hours of observations. These data were reconstructed with both the RTA and the lstchain pipeline. We used the same Monte Carlo simulation to train the random forest used by both pipelines. 

\begin{figure}
    \centering
     \includegraphics[scale=0.6]{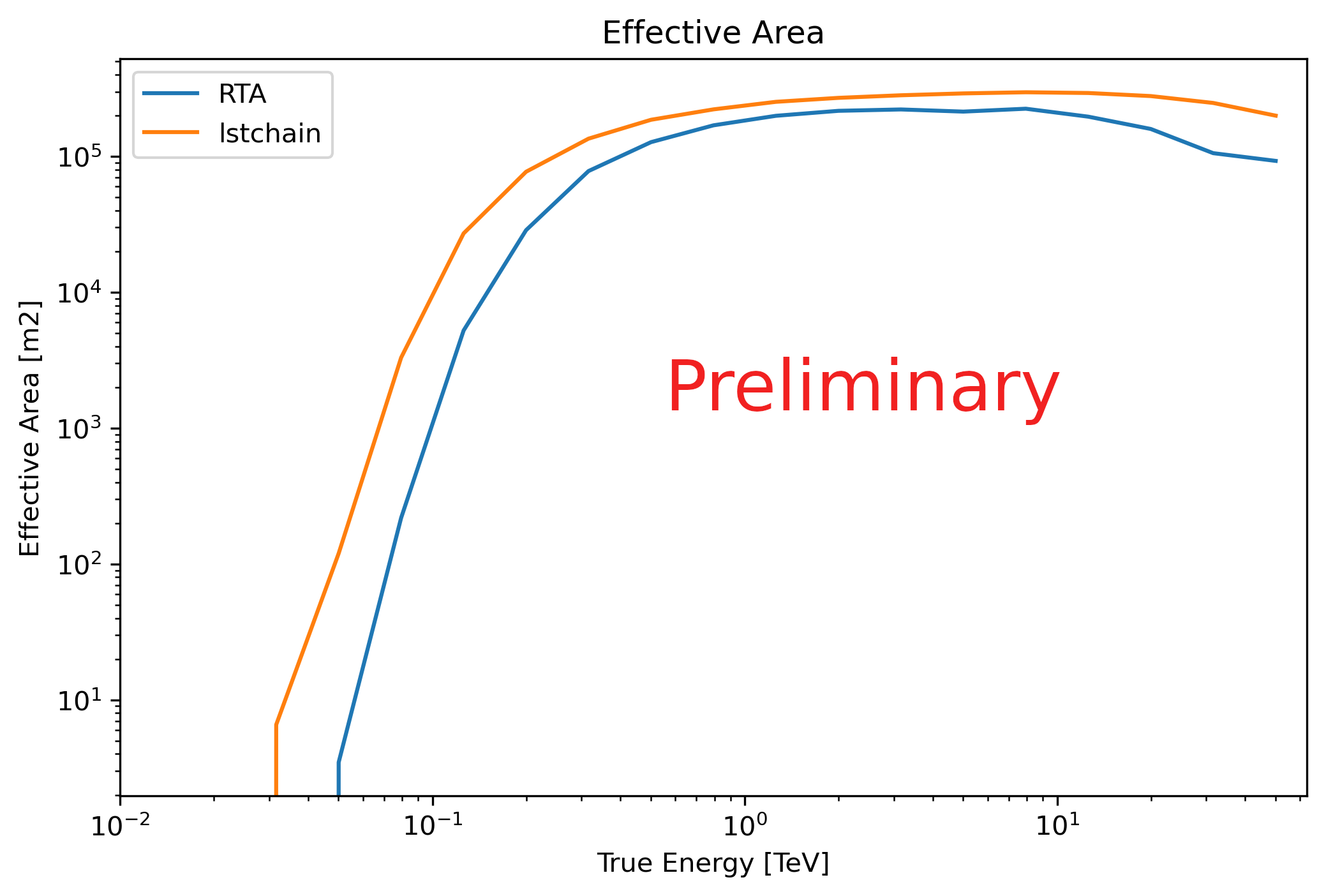}
     \includegraphics[scale=0.5]{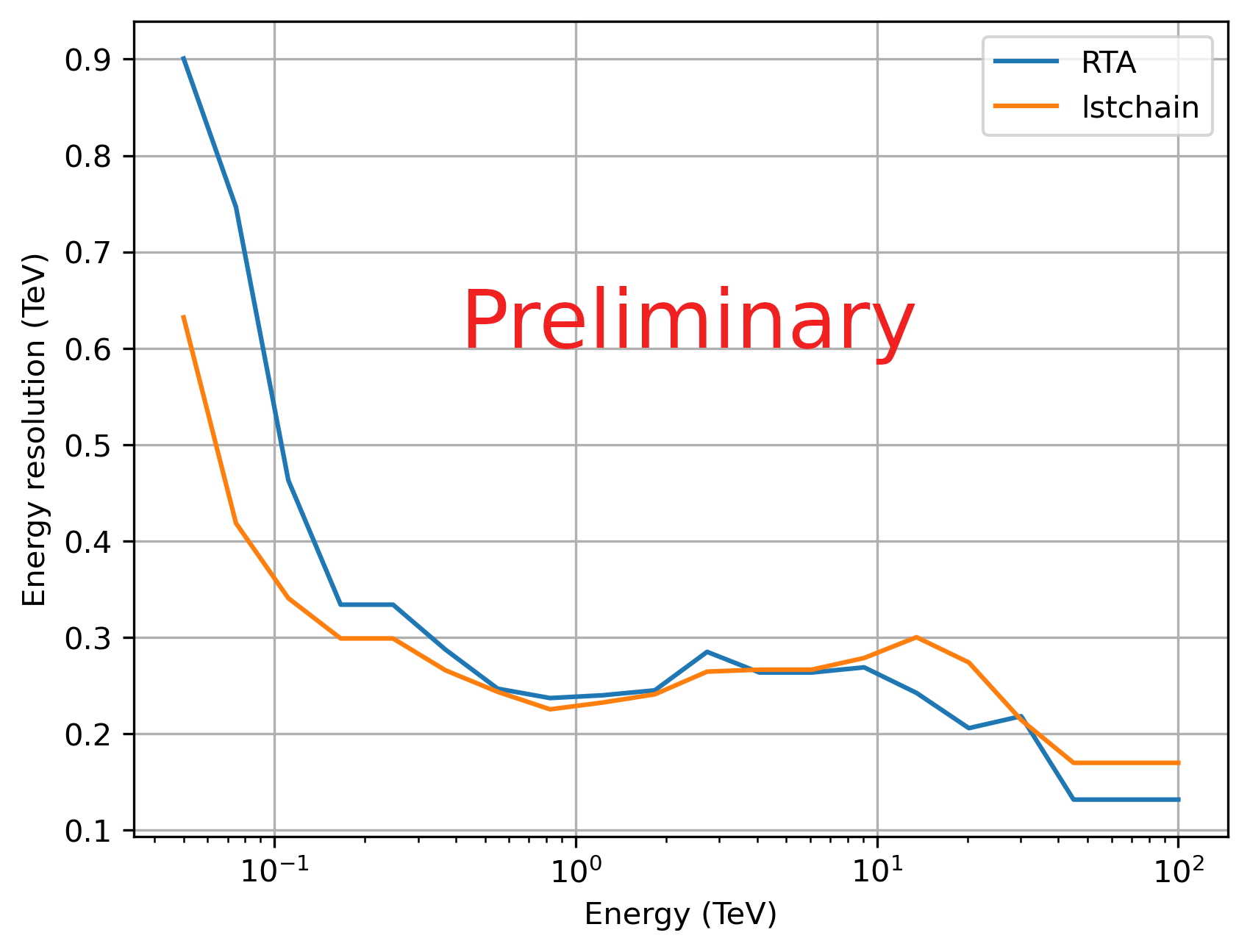}
     \includegraphics[scale=0.5]{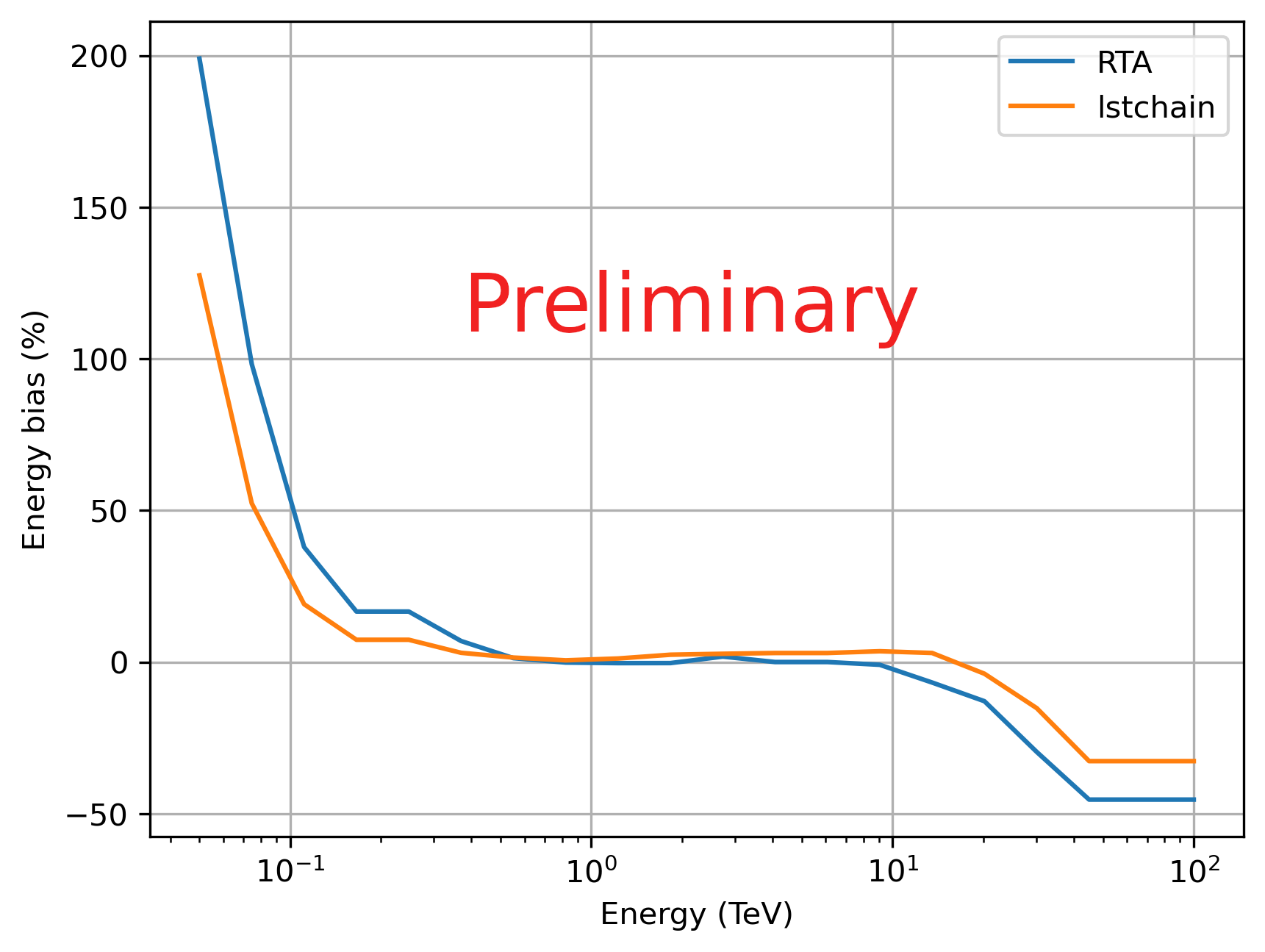}
    \caption{Effective area, energy resolution and bias in function of energy, computed in the gamma Monte Carlo.}
    \label{figRTA}
\end{figure}

The first step consists to optimizing the gammaness threshold. This was done separately on the two chains, by performing a full multiple OFF analysis of a single run of the dataset that is then removed for the rest of the analysis. The size of the ON region is set to $0.2$° and the exclusion region defined as $0.35$°. The same setup will be used in all this proceeding. A maximum significance of $15.3 \sigma$ with a gammaness threshold of $0.75$ for the RTA, and $31.2 \sigma$ with $0.7$ for the offline analysis are found. The effective area, energy bias and resolution, computed using the gamma Monte Carlo simulations, are displayed in figure \ref{figRTA}. The energy resolution and bias are similar between the two analysis chains above $400$ GeV, and a degradation is observed at the lowest energy for the RTA when compared to the lstchain reconstruction. The relative value of the effective area is linked to the leakage of events up to $0.2$° and is thus related to a worse angular resolution for the RTA analysis chain. 

\begin{figure}
    \centering
     \includegraphics[scale=0.42]{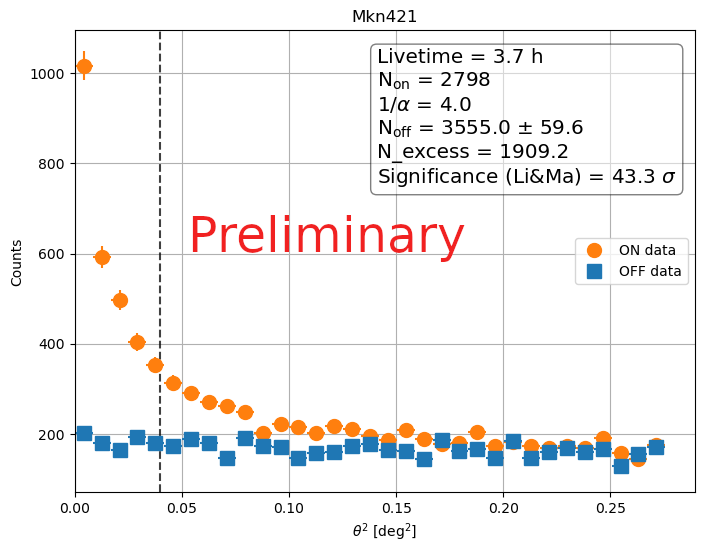}
     \includegraphics[scale=0.42]{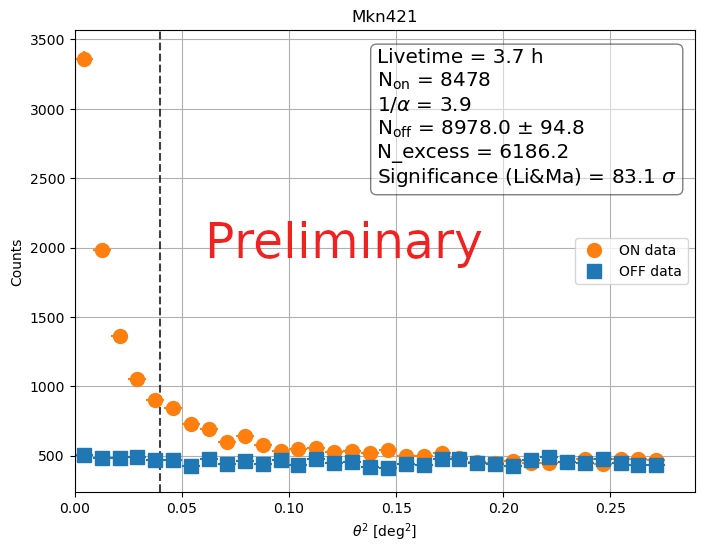}
    \caption{Theta square plot of Markarian 421 data compared between the RTA (left) and the lstchain (right).}
    \label{figRTA_thetasquare}
\end{figure}

A multiple off analysis is performed on the whole dataset, and the resulting theta square plot is presented in figure \ref{figRTA_thetasquare}. A significance of $43.3 \sigma$ for the RTA and $83.1 \sigma$ for the offline analysis are found.

\begin{figure}
    \centering
     \includegraphics[scale=0.51]{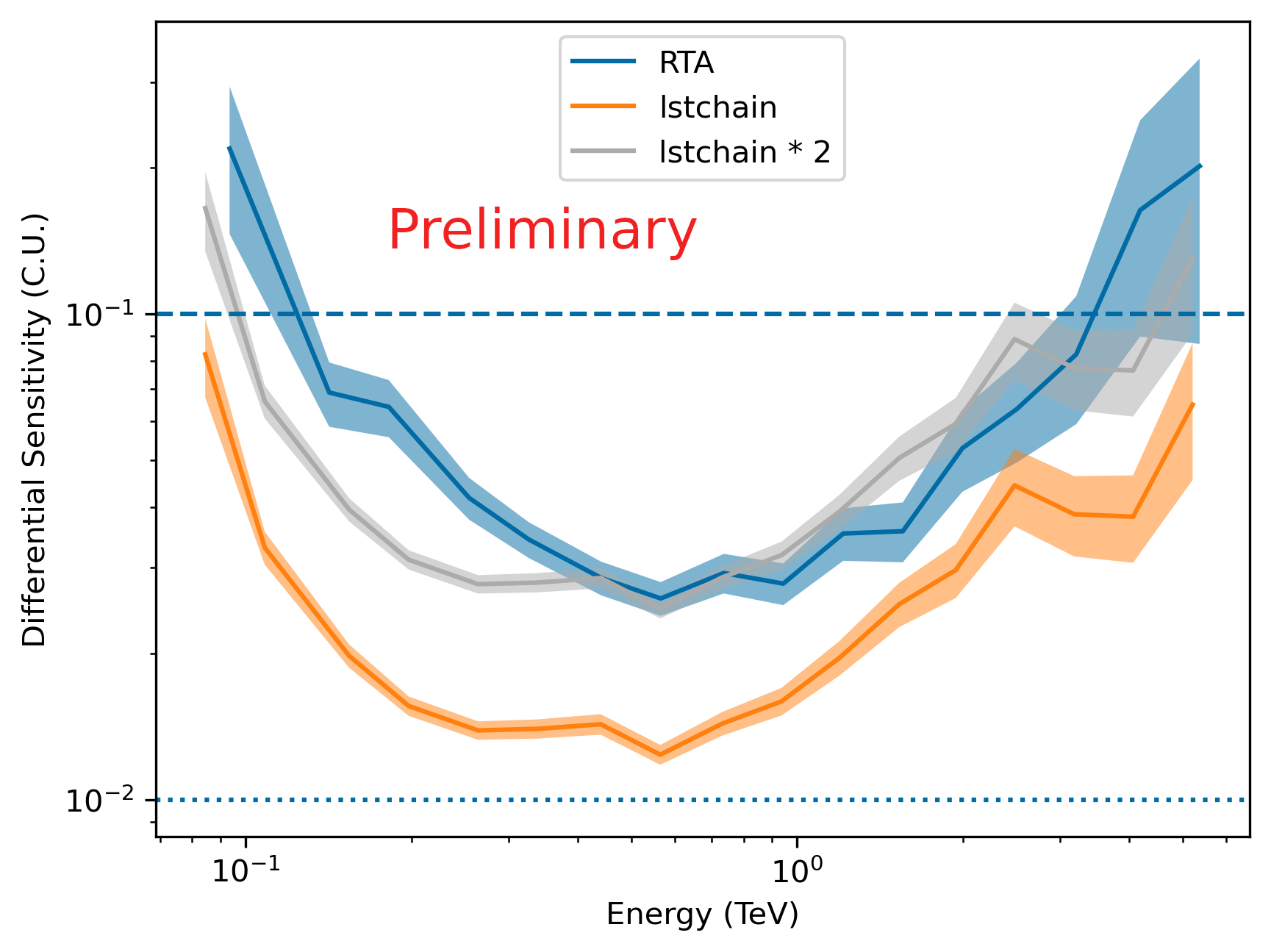}
     \includegraphics[scale=0.51]{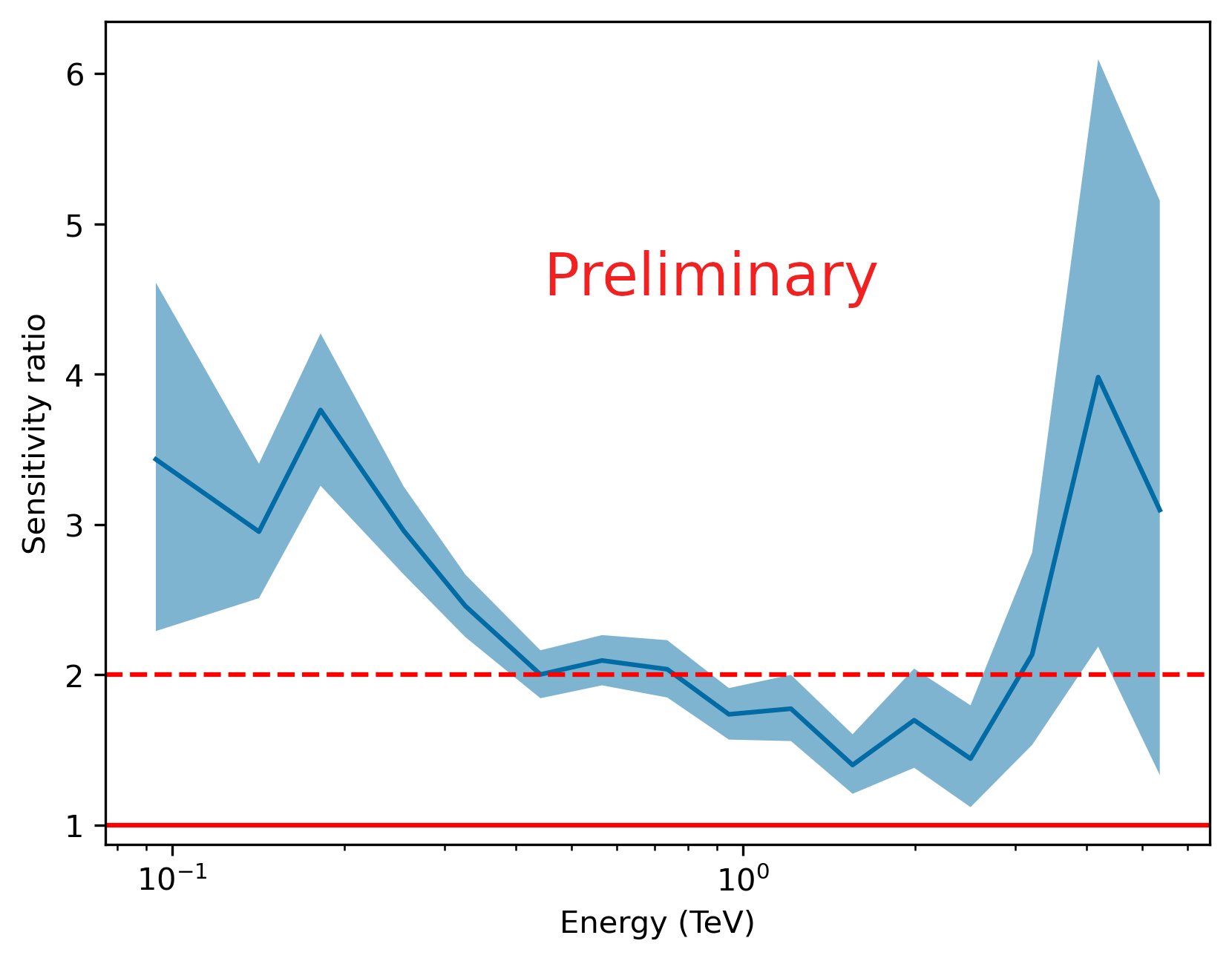}
     \includegraphics[scale=0.51]{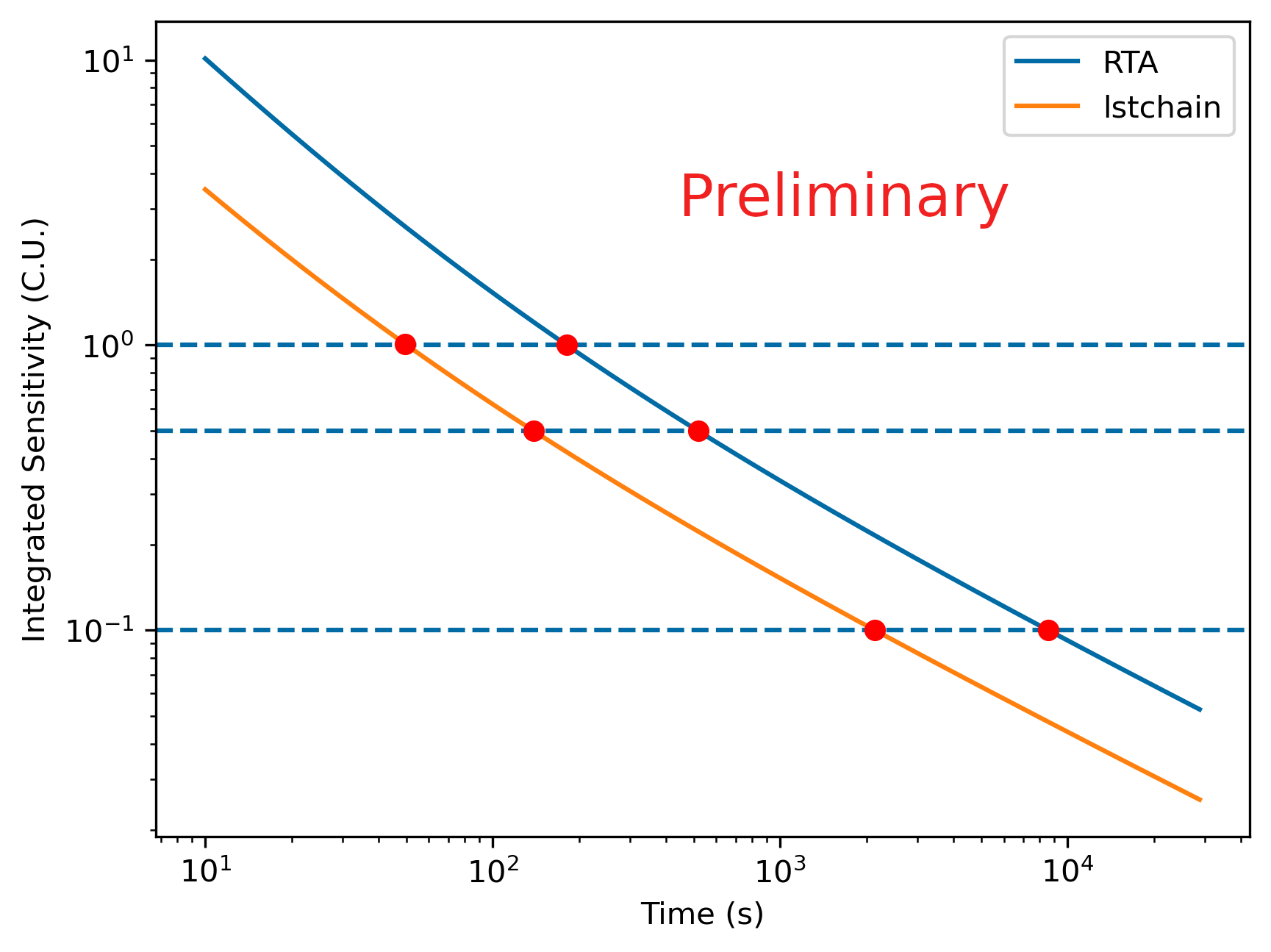}
    \caption{Top left : Differential Sensitivity for 50 hours of the RTA compared with lstchain as a function of true energy. Top right : Sensitivity ratio between the RTA and lstchain as a function of true energy.
    Down : Integral Sensitivity, from $20$ GeV to $10$ TeV, of the RTA compared to a typical flux of $1$, $0.5$ and $0.1$ Crab, as a function of observation time. The time to detect the typical fluxes with the RTA is respectively $177$, $517$ and $8580$ seconds. For the offline analysis, we have respectively $50$, $139$ and $2140$ seconds.}
    \label{figRTA_sensitivity}
\end{figure}

The spectra of the source is reconstructed with the lstchain pipeline, which has been validated for this purpose, and used in order to derive the sensitivity of both chains. Differential sensitivity is defined as the minimal flux needed to have a $5 \sigma$ detection in $50$ hours of observations. The results are presented in figure \ref{figRTA_sensitivity}. The sensitivity of the RTA pipeline is roughly two times worse than the lstchain one above $0.4$ TeV, while at lower energies it worsens.  

Besides the differential sensitivity, an important RTA use case is to detect a source in a time period shorter than a run ($20$ minutes). To explore this possibility, we provide the integral sensitivity versus time, supposing a Crab nebula like spectra, based on the results obtained on the test sample. This is presented in the figure \ref{figRTA_sensitivity}. The RTA is able to detect an integrated flux of $0.3$ Crab units in a single run. 

\section{Conclusions and outlook}

In this proceeding, we presented the software architecture and performance of the RTA of the LST-1. Production of DL3 files from $20$k events takes $66$ seconds, and a flux equivalent to the one of the Crab Nebula above 20 GeV can be detected in $177$ seconds. The differential sensitivity of the RTA is two times worse than the offline analysis chain above $400$ GeV. 

The differences with lstchain is mainly due to two main factors. The first one is the timing parameters computation, particularly useful for the angular resolution, which is performed using a simple linear regression in the RTA for time computing reasons. Moreover, the performed calibration is simplified to gain and pedestal calibration while a full calibration chain imply as well the so-called DRS4 calibration \cite{cite_calib}. %This particular calibration should improve a lot our results particularly at low energy, and will be implemented soon in the received stream by the RTA. 

The RTA will be provided to the Array Control and Data Acquisition System (ACADA) \cite{cite_ACADA} of CTAO for its first release, with no differences with the version presented in this proceeding apart the input calibrated stream. The CTAO speed requirement of $1000$ events/CPU/Tel is already fulfilled. The adaptation of this software to the full array will require modifications mostly at the DL1 level, to merge triggered events, and to DL2 production to take into account the stereoscopic reconstruction. Nevertheless, the scalability and modularity of this software was designed to fulfill this goal in the coming years for the construction of the LST2-4 telescopes and the next releases of ACADA.

\section*{Acknowledgements}
The aknowledgements for CTAO and LST can be found \href{https://www.cta-observatory.org/consortium_acknowledgments/}{here} and \href{https://www.lst1.iac.es/acknowledgements.html}{here}.

\section*{Full Author list of the CTA-LST Project:}

\tiny{\noindent
K. Abe$^{1}$,
S. Abe$^{2}$,
A. Aguasca-Cabot$^{3}$,
I. Agudo$^{4}$,
N. Alvarez Crespo$^{5}$,
L. A. Antonelli$^{6}$,
C. Aramo$^{7}$,
A. Arbet-Engels$^{8}$,
C.  Arcaro$^{9}$,
M.  Artero$^{10}$
% ...
$^{1}$Department of Physics, Tokai University.
$^{2}$Institute for Cosmic Ray Research, University of Tokyo.
$^{3}$Departament de Física Quàntica i Astrofísica, Institut de Ciències del Cosmos, Universitat de Barcelona, IEEC-UB.
$^{4}$Instituto de Astrofísica de Andalucía-CSIC.
$^{5}$EMFTEL department and IPARCOS, Universidad Complutense de Madrid.
$^{6}$INAF - Osservatorio Astronomico di Roma.
$^{7}$INFN Sezione di Napoli.
$^{8}$Max-Planck-Institut für Physik.
$^{9}$INFN Sezione di Padova and Università degli Studi di Padova.
$^{10}$Institut de Fisica d'Altes Energies (IFAE), The Barcelona Institute of Science and Technology.
% ...
}
%% Full authors list (ONLY FOR COLLABORATIONS)
%\clearpage
%\section*{Full Authors List: \Coll\ Collaboration}
%
%\noindent \textbf{Note comment afterwards:} Collaborations have the possibility to provide an authors list in xml format which will be used while generating the DOI entries making the full authors list searchable in databases like Inspire HEP. For instructions please go to icrc2021.desy.de/proceedings or contact us under icrc2021proc@desy.de.\\
%
%\scriptsize
%\noindent
%first.author$^1$, 
%second.author$^2$, 
%third.author$^3$ % .... more names
%and 
%last.author$^{n}$ \\
%
%\noindent
%$^1$first.affiliation.
%$^2$second.affiliation. % .... more affiliation
%$^{m}$last.affiliation.

\end{document}